\documentclass[aps,pra,reprint,showpacs,superscriptaddress,longbibliography]{revtex4-2}
\usepackage{mathtools,amssymb,graphicx,units}

\usepackage[plainpages=false,pdfpagelabels,colorlinks=true,linkcolor=red,urlcolor=blue,citecolor=blue,pdftitle={Title},pdfauthor={},pdfdisplaydoctitle=true,pdfduplex=DuplexFlipLongEdge]{hyperref}
\usepackage{verbatim}
\usepackage[english]{babel}
\usepackage{color}

\usepackage[normalem]{ulem}

\graphicspath{{./figs/}}

\begin{document}

\title{$\pi$-Phase shift across stripes in a charge density wave system}
\author{T. Ying}
\email{taoying86@hit.edu.cn}
\affiliation{School of Physics, Harbin Institute of Technology, Harbin 150001, China}
\author{R.T. Scalettar}
\email{scalettar@physics.ucdavis.edu}
\affiliation{Department of Physics, University of California,
Davis, CA 95616, USA}
\author{R. Mondaini}
\email{rmondaini@csrc.ac.cn}
\affiliation{Beijing Computational Science Research Center, Beijing 100193, China}

\begin{abstract}
Many strongly correlated materials are characterized by
deeply intertwined charge and spin order.
Besides their high superconducting transition temperatures,
one of the central features of these complex patterns in cuprates is a phase shift which occurs 
across lines of decreased hole density.  That is, when doped away from their
AF phase, the additional charge is not distributed uniformly, but rather in 
`stripes'.  The sublattices preferentially occupied by up and down spin are reversed across these
stripes, a phenomonenon referred to as a `$\pi$-phase shift'.
Many of the spin-charge patterns, including the $\pi$-phase shift, are reproduced 
by Density Matrix Renormalization Group and Quantum Monte Carlo calculations of simplified
tight binding (repulsive Hubbard) models.
In this paper we demonstrate that this sublattice reversal is generic by considering
the corresponding phenomenon in the attractive Hubbard Hamiltonian, where a charge density wave phase forms
at half-filling.  We introduce charge stripes via an appropriate local chemical potential;
measurements of charge correlation across the resulting lines of lowered density reveal a clear $\pi$ phase.
\end{abstract}


\maketitle

\section{Introduction}

Inhomogeneous phases are a central feature of strongly correlated materials,
and especially of oxide systems~\cite{dagotto05}.  
The manganites are one example~\cite{dorr06,eerenstein06,israel07,zhang19b}. They have ferromagnetic
and antiferromagnetic (AF) states in close proximity in energy, and when a small
quenched disorder is included, extended glassy regions emerge in which these
phases coexist.  This regime is very sensitive to external perturbation, e.g.~the
application of a small magnetic field, leading to the phenomenon of
colossal magnetoresistance.   The cuprates are another instance.
Here a wide body of experiments, including transport~\cite{doiron07}, 
nuclear magnetic
resonance~\cite{wu11,wu15}, X-ray scattering~\cite{ghiringhelli12,chang12,comin16}, and 
scanning tunneling microscopy~\cite{hoffman02}, has indicated that
complex patterns of charge and spin develop upon 
doping~\cite{tranquada95}, and that these inhomogeneous
structures are also present in the pairing gaps~\cite{julien15,comin16},
thereby suggesting a possible connection to their high superconducting transition 
temperatures.  

Quite remarkably, many of the intricate details of
these structures are reproduced in calculations on simple
Hamiltonians, even when the models are translation invariant~\cite{zheng17}.  
Indeed, one of the earliest indications of stripe physics, 
in which doped holes
arrange themselves in linear patterns, came out of mean-field
calculations~\cite{zaanen89,poilblanc89,Kato90}, pre-dating much of the experimental work.
These observations have been confirmed by a wide variety of methods like functional renormalization group~\cite{yamase16},
the Density Matrix Renormalization Group (DMRG)~\cite{white98,white03,hager05,jiang21}, 
exact diagonalization (ED) ~\cite{fleck00},
dynamic mean field theory~\cite{Vanhala18}, Auxiliary Field
Quantum Monte Carlo (AFQMC)~\cite{chang10,huang17,huang18,qin20}, infinite projected
entangled pair states~\cite{corboz14}, and density matrix embedding theory~\cite{zheng16}, all of which treat many-body effects more exactly.


One central feature of these striped phases in the repulsive Hubbard model is 
their mixed charge and spin character, and, in particular a `$\pi$-phase shift' in the {\it spin} correlations which is found to
exist across a linear depletion of {\it charge}:
The sublattice which holds the surplus of up spin character on
one side of a stripe instead holds a surplus of down spin as the 
stripe is traversed~\cite{qin21}.

In this paper we examine whether such $\pi$-phase shifts exist in a charge
density wave (CDW) phase across a (charge) stripe.
That is, we address the question of whether the mixed character of the
stripe, and the type of order being established across it, is essential
to the existence of a phase shift.  We address this question by
using Determinant Quantum Monte Carlo (DQMC) simulations of the attractive
Hubbard model in which stripes are imposed externally via raising the local
site energy along several rows of the lattice.  While this method does not
establish the spontaneous formation of inhomogeneities, it does allow 
an exploration of sublattice order reversal.
Significantly, because of the absence of a sign problem~\cite{loh90,troyer05}, we
are able to examine low temperatures and especially the effect
of stripes and the $\pi$-phase shift on ($s$-wave) pairing correlations.
Addressing such issues is much more challenging in the repulsive
Hubbard model~\cite{huang17} because of the sign problem.

Our key results are that the introduction of a charge stripe of sufficient
depth does cause the CDW domains on opposite sides of the stripe to
develop a $\pi$-phase shift, so that their high and low density sublattices
are interchanged.  In this way the structure of the charge order
across a density stripe mimics the well-known behavior of the spin 
order in the repulsive Hubbard model.  We also show that this $\pi$-phase shift appears
to be detrimental to pairing order.

This paper is organized as follows.  In Sec.~\ref{sec:ham_meth} we define the Hamiltonian and correlation
functions we will investigate, and give a brief description of the DQMC and ED methodologies.  Section \ref{sec:res} presents our results for charge and pairing correlations,
focusing especially on the issue of a $\pi$-phase shift. Finally,
Sec.~\ref{sec:concl} summarizes our conclusions.

\section{Hamiltonian and methodology} \label{sec:ham_meth}
\begin{figure}[t!] 
\includegraphics[width=0.9\columnwidth]{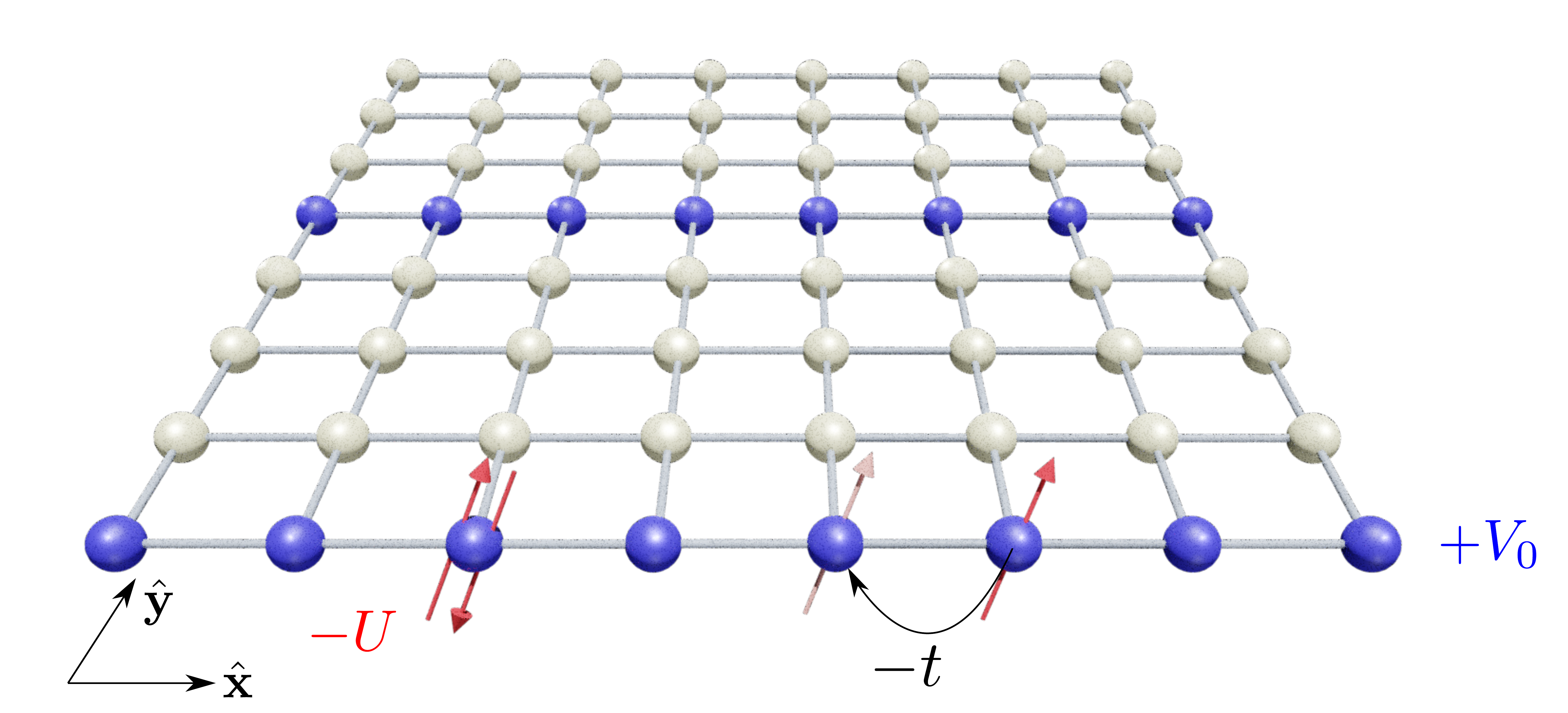}
\caption{(Color online) Illustration of the Hamiltonian in an $8\times 8$ lattice with period ${\cal P}=4$; sites with $V_0$ active are depicted in blue, whereas the interstripe sites are white. The other relevant parameters, hopping and attractive interactions are annotated.
\label{fig:stripesketch}
}
\end{figure}

We study a two dimensional square lattice attractive Hubbard Hamiltonian
in which stripes are introduced externally
via a raised site energy $V_0$ on a set of rows of period ${\cal P}$, 
${\bf i}=(i_x,i_y)$ with ${\rm mod}(i_y,{\cal P})=0$. 
\begin{align}
\hat {\mathcal H}=-t\sum_{\langle {\bf i\,j} \rangle \, \sigma}
(\hat c_{{\bf i}\sigma}^\dagger \hat c_{{\bf j}\sigma}^{\phantom{\dagger}} +
\hat c_{{\bf j}\sigma}^\dagger \hat c_{{\bf i}\sigma}^{\phantom{\dagger}} )
+ U \sum_{\bf i} \hat n_{{\bf i}\uparrow} \hat n_{{\bf i} \downarrow}
\nonumber \\
-\mu \sum_{{\bf i}}
(\hat n_{{\bf i}\uparrow} + \hat n_{{\bf i}\downarrow})
+V_0 \sum_{i_y \in {\cal P}}
(\hat n_{{\bf i}\uparrow} + \hat n_{{\bf i}\downarrow}). 
\label{eq:ham}
\end{align}
We choose ${\cal P}=4$,
so that each stripe (row with $V_0$ term active, blue spheres in Fig.~\ref{fig:stripesketch})
is separated by three rows where the $V_0$ term
is not present (light-colored spheres in Fig.~\ref{fig:stripesketch}).

Our primary methodology is the DQMC approach~\cite{Blankenbecler81,white89a}.  
We first express the partition function
associated with the Hamiltonian of Eq.~\ref{eq:ham},
${\cal Z} = {\rm Tr} \, e^{-\beta \hat{\cal H}}$, as a path integral by
discretizing the imaginary time $\beta = L_\tau \Delta \tau$.
This allows us to use the Trotter approximation
$e^{-\Delta \tau \hat {\cal H}} \approx
e^{-\Delta \tau \hat {\cal K}} 
e^{-\Delta \tau \hat {\cal V}} $
where $\hat {\cal K}$ and $\hat {\cal V}$ are the single- and two-particle
terms of $\hat {\cal H}$ respectively.  
In this work we use $\Delta \tau$ = 0.1; an analysis of Trotter
errors in observables is given in Appendix \ref{sec:app_Trotter}.
With $\hat {\cal V}$ isolated, we
can introduce a (discrete) Hubbard-Stratonovich field (HSF)~\cite{hirsch89a} to decouple
the interactions.  The trace which gives ${\cal Z}$ now contains
only exponentials of quadratic forms of fermion operators and can be done
analytically.  The resulting expression for ${\cal Z}$ is a sum
over the configurations of the HSF, which is done stochastically.  We will present results mainly for
$8 \times 8$ ($12 \times 12$ and $16 \times 16$-lattice results are shown in Appendix \ref{sec:app_fse_fte}), i.e.~linear extent $L=8$, and
supplement the DQMC calculations with ED to
confirm and further understand the physics.  Our ED calculations, presented in the
Appendix \ref{sec:app_ed}, 
focus on a $2 \, \times \,4$ lattice, which has a manageable
Hilbert space size yet is large enough to hold a stripe and examine
the spin and charge patterns at finite temperature $T$.

For $V_0 \neq 0$, the density is inhomogeneous.  We denote by $\rho$ the overall density,
averaged over the entire lattice.  $\rho_{\rm str}$ is the density on the stripes, the sites with $V_0 \neq 0$, and
$\rho_{\rm dom}$ is the density on the domains between the stripes.  For ${\cal P}=4$ these are related by
$\rho = \frac{1}{4} \rho_{\rm str} + \frac{3}{4} \rho_{\rm dom} $.


We focus attention on the equal-time, real space density-density correlation function
\begin{align}
c({\bf r}) 
=\big\langle \,(\hat n_{\bf i} - 1) (\hat n_{\bf i+r} - 1) \, \big\rangle
\end{align}
between two sites ${\bf i}$ and ${\bf j}$.  
The subtraction is a convention, useful for this paper where much of the
lattice has $\langle \hat n_{\bf i} \rangle = 1$, since it then enables a focus on
fluctuations away from the average density.
In a homogeneous CDW phase,
this correlation function is long-ranged, taking positive values
for pairs of sites on the same sublattice and negative values for sites
on opposite sublattices.  We also examine the real space, equal-time 
$s$-wave pairing correlators $p({\bf r})$, and their associated structure factor $P_s$ at ${\bf q}=0$:
\begin{align}
p({\bf r})  &= 
\big\langle \, \hat \Delta^{\phantom{\dagger}}_{\bf i+r} 
\hat \Delta^{\dagger}_{\bf i} \, \big\rangle
\hskip0.35in 
\hat \Delta^{\dagger}_{\bf i} \equiv 
\hat c^{\dagger}_{{\bf i}\uparrow}
\hat c^{\dagger}_{{\bf i}\downarrow}
\nonumber \\
P_s &= \sum_{\bf r} \, p({\bf r})
\end{align}
which signal the formation of a superconducting
phase.  In the presence of off-diagonal long range order
$p({\bf r})$ approaches a non-zero value as ${\bf r} \rightarrow \infty$,
and $P_s$ grows linearly with the lattice size $L^2$.

In the absence of $V_0$, and at half-filling
$\mu = U/2$, the attractive Hubbard model has simultaneous CDW and $s$-wave 
order.  This enlarged symmetry of possible ordered phases implies that long range
order (LRO) is possible only at $T=0$ \cite{scalettar89,moreo91}.
On finite lattices, LRO can be observed as long as the temperature is
lowered to a value for which the correlation length $\xi$
exceeds the linear lattice size $L$.
The introduction of stripes ($V_0 \neq 0$) breaks this
degeneracy between CDW and pairing.
As we will discuss below, the resulting anisotropy favors pairing order, if small in strength,
an observation which will be confirmed by our DQMC simulations at small values of $V_0$.
The qualitative argument is similar to that for uniform doping
of the attractive Hubbard model~\cite{scalettar89,moreo91}.


\section{Results} \label{sec:res}

We first focus separately on the density and pairing correlations,
and then comment on how they are related, interpreting the enhancement
of pairing order for weak stripes in terms of a particle-hole transformation
to the repulsive Hubbard model.

\subsection{Density-density correlations: Demonstration of Stripe Formation}

\begin{figure}[t!] 
\includegraphics[width=0.99\columnwidth]{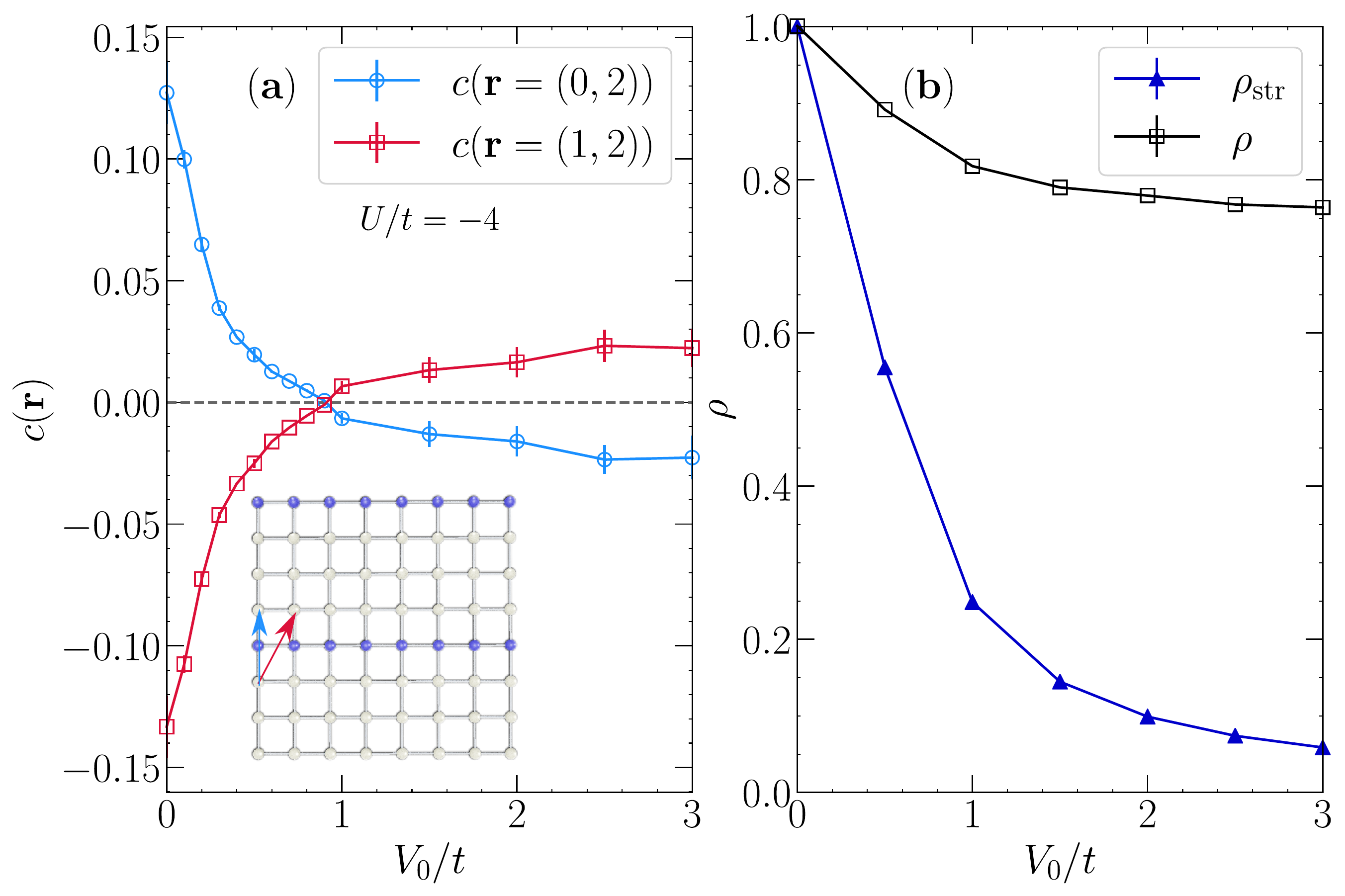}
\caption{(a) Density-density correlator 
$c\big({\bf r} = (0,2)\big)$
(blue circles and
blue arrow in the inset) and 
$c\big({\bf r} = (1,2)\big)$
(red squares, and red arrow in the inset), 
as functions of stripe strength $V_0$. The emergence of $\pi$-phase shift at $V_0 \sim t$ is signalled by the reversal in sign of the correlation functions. (b) The total electron density $\rho$ and the density on the stripe $\rho_{\rm str}$, as functions of stripe strength $V_0$. The electron density off stripe $\rho_{\rm dom}$ is fixed to be $1$ by adjusting the global chemical potential $\mu$. Here the lattice size is $8 \, \times \, 8$, the on-site attraction $U = -4\,t$ is half the non-interacting bandwidth, and the inverse temperature $\beta \,t = 10$.
\label{corr_cross_rhooff1}
}
\end{figure}

We begin with DQMC results for the density-density correlations.
The central feature of the $\pi$-phase shift is the interchange of the sublattice
ordering pattern
across the stripe.  A spin (or charge) correlation connecting sites on the
{\it same} sublattice,
which would be positive in the absence of a stripe, becomes negative.  
Likewise, 
a spin (or charge) correlation connecting sites on the
{\it different} sublattices,
which would be negative in the absence of a stripe, becomes positive.  
The pairs of sites connected by blue and red arrows in the inset of 
Fig.~\ref{corr_cross_rhooff1}(a) show these two cases.

We perform a set of simulations in which $V_0$ is increased, gradually forming
well-defined stripes of reduced fermion occupation.  We simultaneously adjust the global
chemical potential $\mu$ to keep the density on the sites
of the domains between the stripes $\rho_{\rm dom}=1$,
a value which is optimal for CDW order (other densities are investigated in Appendix \ref{sec:app_other_dens}).
Figure \ref{corr_cross_rhooff1}(a) is one of the central results of our paper. It demonstrates that as stripes are introduced, a $\pi$-phase shift emerges.  For the parameters shown, $U=-4\,t$ and $T=t/10$, the crossover occurs at $V_0 \sim t$.

 
The corresponding density evolution is shown in Fig.~\ref{corr_cross_rhooff1}(b). The blue curve shows the rapid depletion of fermion occupation on the stripes.  At $V_0 \sim t$ where the $\pi$ phase shift emerges, $\rho_{\rm str} \sim 0.25$. The black curve shows the overall fermion occupation $\rho$, which approaches 0.75 in the large $V_0$ limit since 1/4 of the sites (the stripes) have been driven to empty with the remaining 3/4 of the sites remaining at unit density. The two curves are not independent, being related, as noted earlier, by  $\rho = \frac{3}{4} + \frac{1}{4} \rho_{\rm str}$; both are shown for clarity, however.

If one instead investigates the \textit{repulsive} Hubbard model ($U>0$), see Appendix \ref{sec:app_plusU}, a qualitatively similar picture follows: Sufficiently large stripe energies lead to the emergence of a \textit{magnetization} reversal, i.e., a magnetic $\pi$-phase shift appears, whose onset, however, occurs at much larger $V_0$'s ($V_0 \simeq 4t$ for $U=+4t$).

\begin{figure}[h] 
\includegraphics[width=0.99\columnwidth]{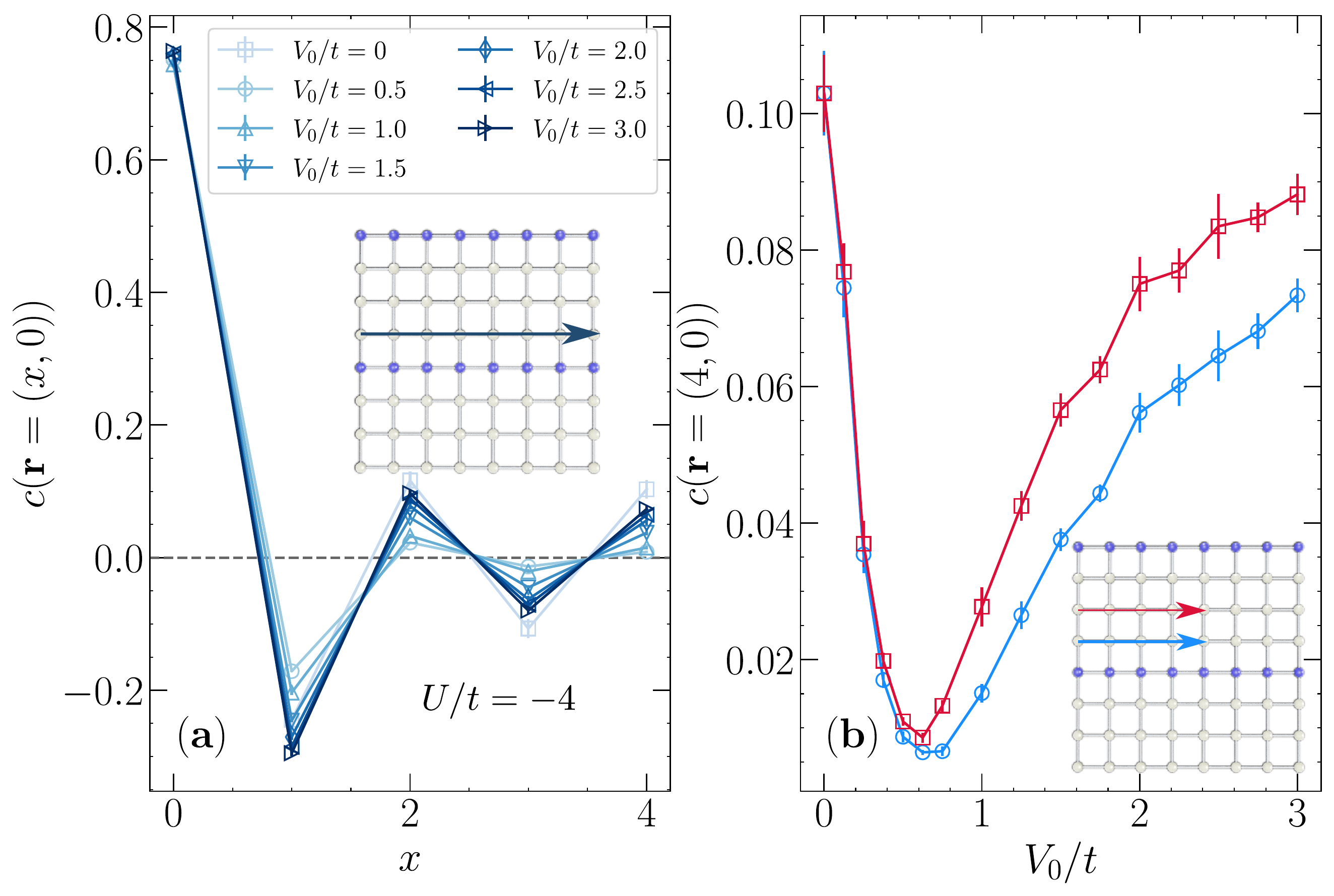}
\caption{ (a) Density-density correlations $c\big({\bf r}=(x,0)\big)$ on a row parallel to, and immediately neighboring, the stripe, as indicated in the lattice of the inset. Results for different values of the stripe strength $V_0$ are shown. (b) Density-density correlations 
$c\big(\,{\bf r}=(4,0)\,\big)$ on a row parallel to the stripe, for both immediately neighboring and in the middle of two stripes, as indicated in the lattice of the inset. Results for different values of the stripe strength $V_0$ are shown. In both panels, the density in the inter-stripe domains is fixed at $\rho_{\rm dom}=1$ by adjusting $\mu$. Here the lattice size is $8 \, \times \, 8$, the on-site attraction $U = -4\,t$, and the inverse temperature $\beta \,t = 10$.
\label{C(r)-NeighborStripe-rhooff1}
}
\end{figure}

Turning back to the attractive case, to characterize the nature of the charge pattern more precisely,
Fig.~\ref{C(r)-NeighborStripe-rhooff1}(a) shows
the density-density correlation function $c\big(\,{\bf r}=(x,0)\,\big)$ 
between sites running
immediately parallel to the charge stripe.
These correlations exhibit an interesting non-monotonicity with $V_0$.
As might be expected, the oscillating CDW pattern is most robust at $V_0=0$
when one has a pristine half-filled lattice.  For weak stripe potentials
$V_0 = 0.5, 1.0$, there is only short range order,
and $c\big(\,{\bf r}=(4,0)\,\big) \sim 0$.
However, as $V_0$ is further increased, CDW order is recovered. 
For $V_0=3$, $c\big(\,{\bf r}=(4,0)\,\big)$
is nearly as large as its $V_0=0$ value.
 Fig.~\ref{C(r)-NeighborStripe-rhooff1}(b) emphasizes this behavior by showing
$c\big(\,{\bf r}=(4,0)\,\big)$ as a function of $V_0$. As a comparison, we also show $c\big(\,{\bf r}=(4,0)\,\big)$ along the middle line in the domain as a function of $V_0$.


\begin{figure}[t!] 
\includegraphics[width=1\columnwidth]{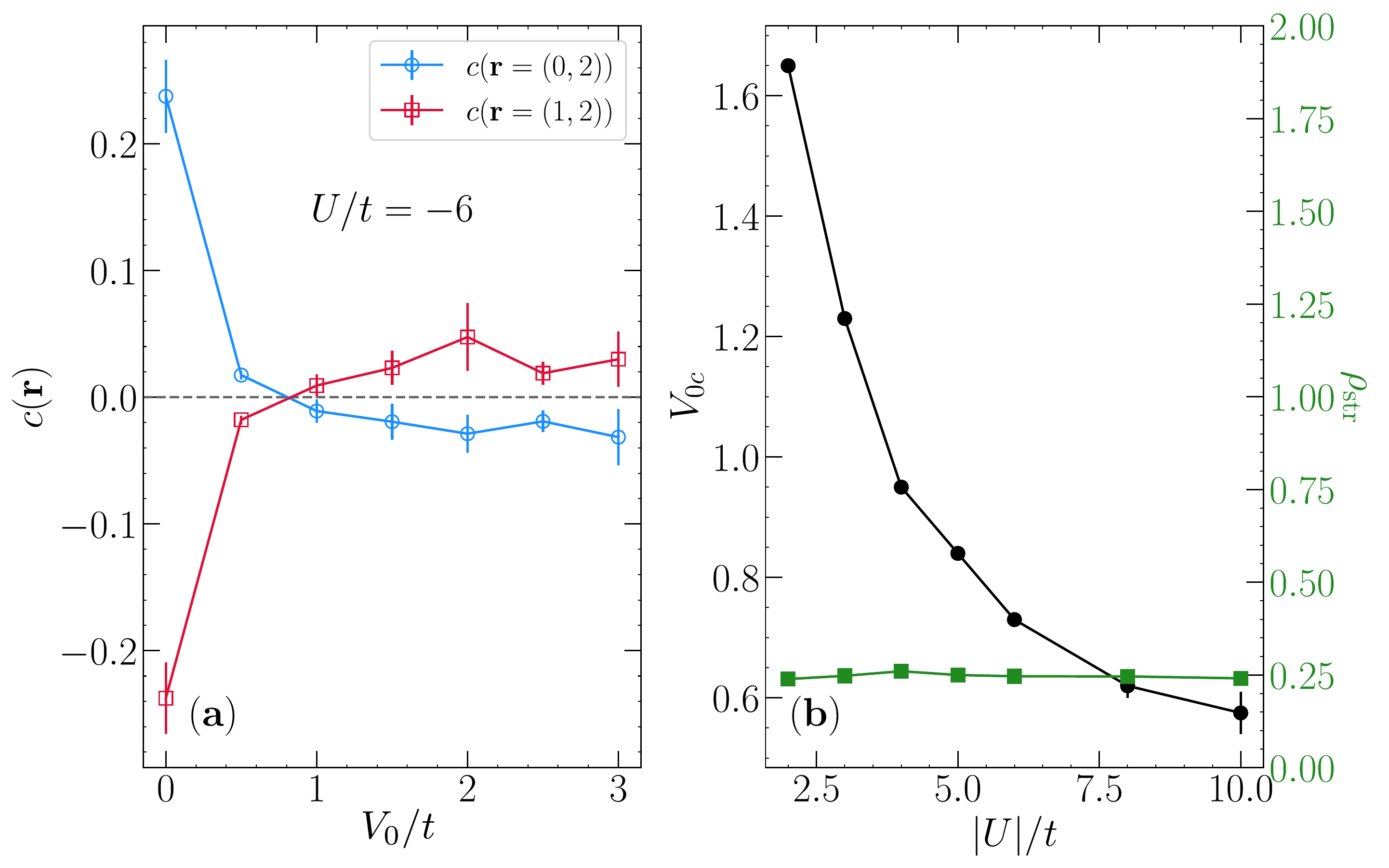}
\caption{(a) Similar to Fig.~\ref{corr_cross_rhooff1} (a), but with the interaction strength increased to $U=-6\,t$. The formation of $\pi$-phase shift occurs at a slightly smaller stripe potential.
(b) Compilation of the critical values $V_{0c}$ where the $\pi$-phase shift appears with different interaction strengths, and the corresponding electron density on the stripe $\rho_{\rm str}$. The inverse temperature is set at $\beta t=10$.
\label{corr_cross_rhooff1-U-6}
}
\end{figure}

The preceding results are for a coupling $U=-4\,t$ which is one-half the
non-interacting bandwidth.  The low temperature spin and charge correlations
of the Hubbard Hamiltonian
are usually strongest at somewhat larger interaction strengths.
In Fig.~\ref{corr_cross_rhooff1-U-6}(a) we examine $c({\bf r})$ 
across a stripe for $U=-6\,t$.
The $\pi$-phase shift forms at slightly smaller values of $V_0$.
A compilation of the critical energies $V_0$ that trigger a $\pi$-phase shift as a function of $U$ is given in Fig.~\ref{corr_cross_rhooff1-U-6}(b), showing both the critical $V_{0c}$, and the corresponding fermion occupation on the stripe $\rho_{\rm str}$. The trend above described is confirmed, that a smaller stripe energy $V_0$ with systematically larger interactions (in magnitude) is sufficient for the onset of a $\pi$-phase shift. Even more remarkably, the electronic density at the stripes where the transition occurs is largely $U$-independent in the range investigated, pinned at filling $\rho_{\rm str}=0.25$. This corresponds to the total density $\rho=0.8125$ [$\rho_{\rm dom} = 1$]. 

In the Appendix \ref{sec:app_fse_fte} we show that these results for the cross-stripe spin correlations are converged in both inverse temperature $\beta$ and spatial lattice size $L$.

\subsection{$s$-wave pairing correlations}
\begin{figure}[t!] 
\includegraphics[width=1\columnwidth]{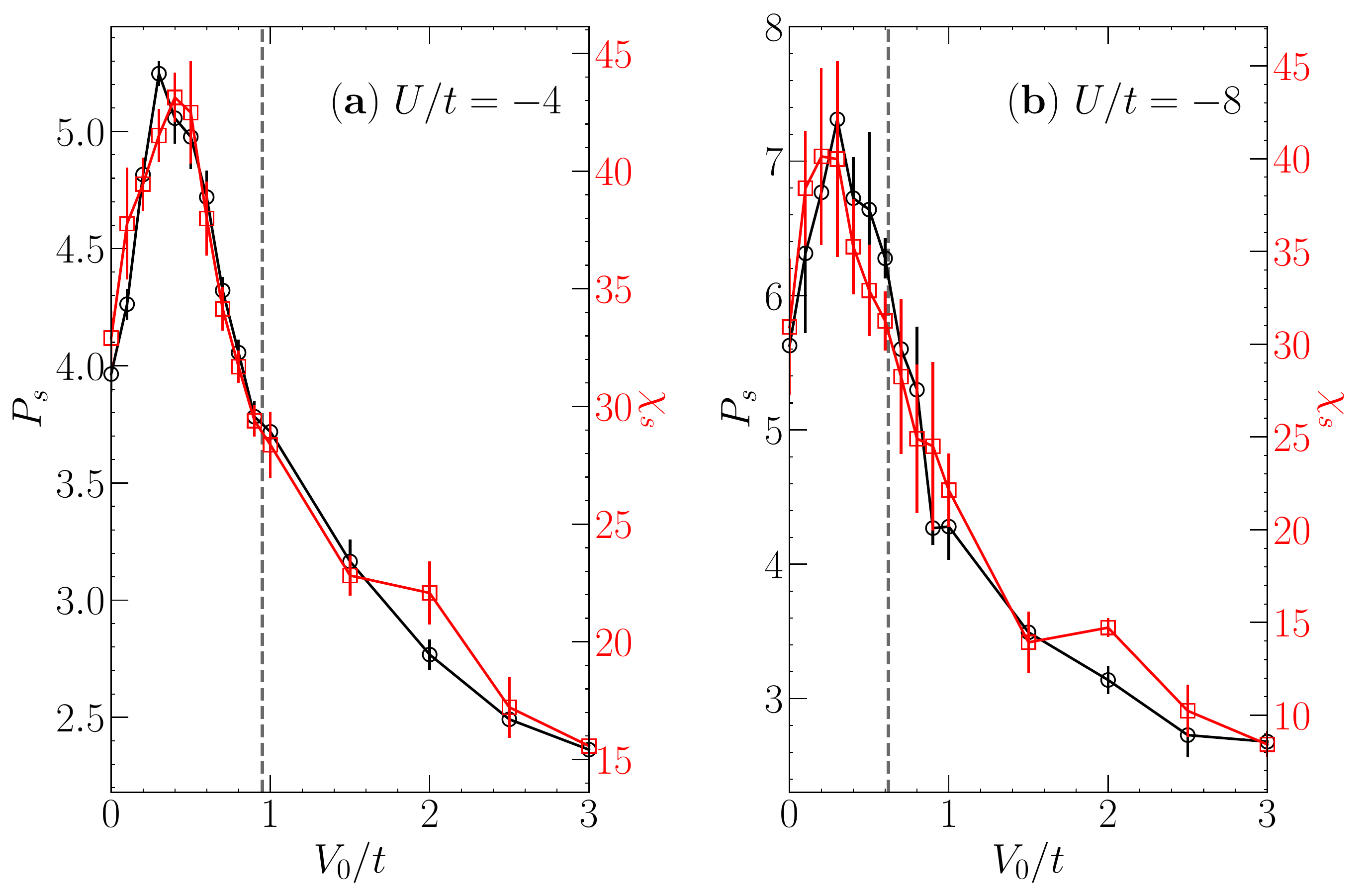}
\caption{The $s$-wave pairing structure factor $P_s$ (left y-axes) as a function of $V_0$ and the corresponding pair susceptibility $\chi_s$  (right y-axes) for $\rho_{\rm dom}=1$, in an $8\times 8$ lattice with $U/t = -4$ (a) and $-8$ (b). Apart from the very different scales, the two quantities qualitatively follow each other. Vertical dashed lines give the critical stripe strength for $\pi$-phase shift formation $V_{0c}(U)$. The inverse temperature $\beta t = 10$.
\label{Ps_chi_s}
}
\end{figure}

We next look at how the $s$-wave pairing is affected by the stripe strength. We begin with the pair structure factor $P_s$, focusing, as before, on fermionic `domain' occupation $\rho_{\rm dom}=1$. Figure~\ref{Ps_chi_s} shows the dependence of $P_s$ on $V_0$. Initially (small
$V_0$), the presence of an extrinsic density modulation results in enhanced $s$-wave pairing. This increase exists roughly up to the
value of $V_0=V_{0c}(U)$ where the charge $\pi$-phase shift is formed. Subsequently, a further increase in the stripe strength is detrimental to the pairing. This behavior is consistent with trends at different values of the interaction $U$ [See Figs.~\ref{Ps_chi_s}(a) and \ref{Ps_chi_s}(b)]. 

The pair-field susceptibility,
\begin{align}
\chi_s = \sum_{\bf r} \int d\tau  \,\,
\langle e^{\tau \hat {\cal H}}\, \hat \Delta_{\bf i+r}\, e^{-\tau \hat {\cal H}} \, \hat \Delta^{\dagger}_{\bf i}\rangle,
\label{eq:chis}
\end{align}
provides a more sensitive measurement
of superconductivity (SC), since it also samples over the correlations in imaginary time.
$\chi_s$ as a function of $V_0$ for the density $\rho_{\rm dom}=1$ is shown 
(red line) in Fig.~\ref{Ps_chi_s}. The pair structure factor $P_s$ (black line of Fig.~\ref{Ps_chi_s}) and $\chi_s$ maintain the same ratio
independent of $V_0$: $\chi_s/P_s \sim \beta$.
Indeed, since $\chi_s$ includes $\int_0^\beta d\tau$ (see
Eq.~\ref{eq:chis}), a factor of $\beta$ is the
`expected' ratio, assuming there is very little decay in imaginary time.
The data of 
Fig.~\ref{Ps_chi_s} therefore are indicative of long range pair
correlations in $\tau$.

\subsection{Intertwined Order}

The preceding results suggest that in this model pairing is initially enhanced when charge stripes are introduced, but that the subsequent development of a $\pi$-phase shift between the CDW domains then rapidly suppresses the SC. This is a non-trivial result.  One might intuitively expect that turning on $V_0$ would promote SC by doping the stripes away from half-filling, but the subsequent diminution by the emergence of CDW sublattice reversal does not have an obvious origin.

We can get a more nuanced view of the interplay of the charge patterns
and pairing by recalling that,
on a bipartite lattice,
the Hamiltonian of Eq.~\ref{eq:ham} with $U<0$ has a well known mapping 
which reverses the sign to $U>0$.  This is accomplished via a partial particle-hole
transformation (PHT) on just one of the spin species:
$c_{{\bf i} \uparrow} \rightarrow (-1)^{\bf i} c_{{\bf i} \uparrow}$, leaving
$c_{{\bf i} \downarrow}$ unchanged.
Here  $(-1)^{\bf i} = +1(-1)$ on sublattice {\cal A}({\cal B}).
The SC correlations in the attractive case map onto AF correlations in the $xy$
plane of the repulsive model, while CDW correlations map onto AF correlations in the $z$ direction of the repulsive model.

Under this PHT,
the additional on-site
term $V_0$, and global chemical potential $\mu$, couple to the $z$ component of spin
$S^z_{\bf i}= n_{{\bf i} \uparrow} - n_{{\bf i} \downarrow}$. 
As has previously been noted~\cite{scalettar89,moreo91}, the full symmetry of the spin correlators of 
a Heisenberg antiferromagnet is broken in a Zeeman field.   $B_z$ will preferentially favor
order in the $xy$ plane.  Returning to the $-U$ model, we conclude that the $V_0$ and
$\mu$ terms are both likely to favor SC over CDW formation, as we indeed observe for $V_0 \lesssim t$.
This analysis, while very useful in the $V_0=0$ case, does not appear to
lend additional insight into why the enhancement of SC is limited to stripes which are not
accompanied by a $\pi$-phase shift.

Finally, note that
this familiar mapping of attractive to repulsive Hubbard model
does {\it not} allow one to infer stripe formation in the attractive case from
existing results for $U>0$.
(A full PHT preserves the sign of $U$, and instead relates the model at 
fixed $U$ and $V_0$ to the model at the same $U$ but reversed
$-V_0$.)

\section{Conclusions} \label{sec:concl}

In this paper we have shown that the well-known behavior of the AF
spin order across a charge stripe in the repulsive Hubbard model, in which the
magnetization of the sublattices on opposite sides of a stripe
is interchanged, also occurs
in CDW order in the attractive Hubbard model.  
Thus the fact that the magnetic order in the inter-stripe domain is distinct 
from the charge pattern of the stripe appears to be irrelevant to the
occurrence of a $\pi$-phase shift.  That phenomenon occurs even if the 
inter-stripe domain order and stripe physics are both associated with charge
degrees of freedom.
We note that 
in this work we have exclusively consider stripes which are imposed externally
via the potential $V_0$.  
We have not considered spontaneous stripe formation, which
is a considerably more challenging calculation.  
In the much-studied repulsive Hubbard mode ($U>0)$ case, 
spontaneous stripes were found early on in inhomogeneous Hartree-Fock
calculations~\cite{zaanen89,poilblanc89}, but are very challenging to
see in DQMC~\cite{huang17,huang18}.  

Although we have found here an {\it analogous} $\pi$-phase shift
between CDW across a stripe in the attractive Hubbard model, similar
to what is known for SDW domains in the repulsive case, there is
also a potential {\it difference} between the two situations.
DQMC studies of the $+U$ Hamiltonian~\cite{mondaini12} 
found an increase of the stripe strength and the $\pi$-phase shift
keeps enhancing pairing monotonically.  In contrast, we find here
that the evolution of superconducting correlations with stripe stength is
non-monotonic, and that when the stripes are sufficiently robust
to support a $\pi$-phase shift, pairing is suppressed.  
A DCA treatment~\cite{maier10} of  the repulsive case 
indicates an optimal stripe strength for pairing, but did not examine the $\pi$ phase shift.

\section{Acknowledgments}
T.Y.~was supported by the joint guiding project of Natural Science Foundation of Heilongjiang Province (Grant No.~LH2019A011).
R.T.S.~was supported by the grant DE‐SC0014671 funded by the U.S. Department of Energy, Office of Science.
R.M.~acknowledges support from the National Natural Science Foundation of China (NSFC) Grants No.~U1930402, 12050410263, 12111530010 and No.~11974039. 

\section{Appendix}
These Appendices are divided as follows. Appendix \ref{sec:app_plusU} contrasts some of the results to the case with repulsive interactions. Appendices \ref{sec:app_Trotter} and \ref{sec:app_fse_fte} consider the systematic Trotter, finite temperature, and finite lattice size errors in our DQMC calculations, showing that they do not affect our conclusions. Appendix \ref{sec:app_other_dens} analyzes the charge patterns when $\mu$ is not tuned to keep the density in the regions between the stripes at the commensurate filling which is optimal for CDW, $\rho_{\rm dom}=1$. Finally, Appendix \ref{sec:app_ed} compares our DQMC results with ED, showing that the qualitative physics is unchanged.
\appendix 

\section{The repulsive case: magnetic $\pi$-phase shift} \label{sec:app_plusU}

\begin{figure}[h] 
\includegraphics[width=0.99\columnwidth]{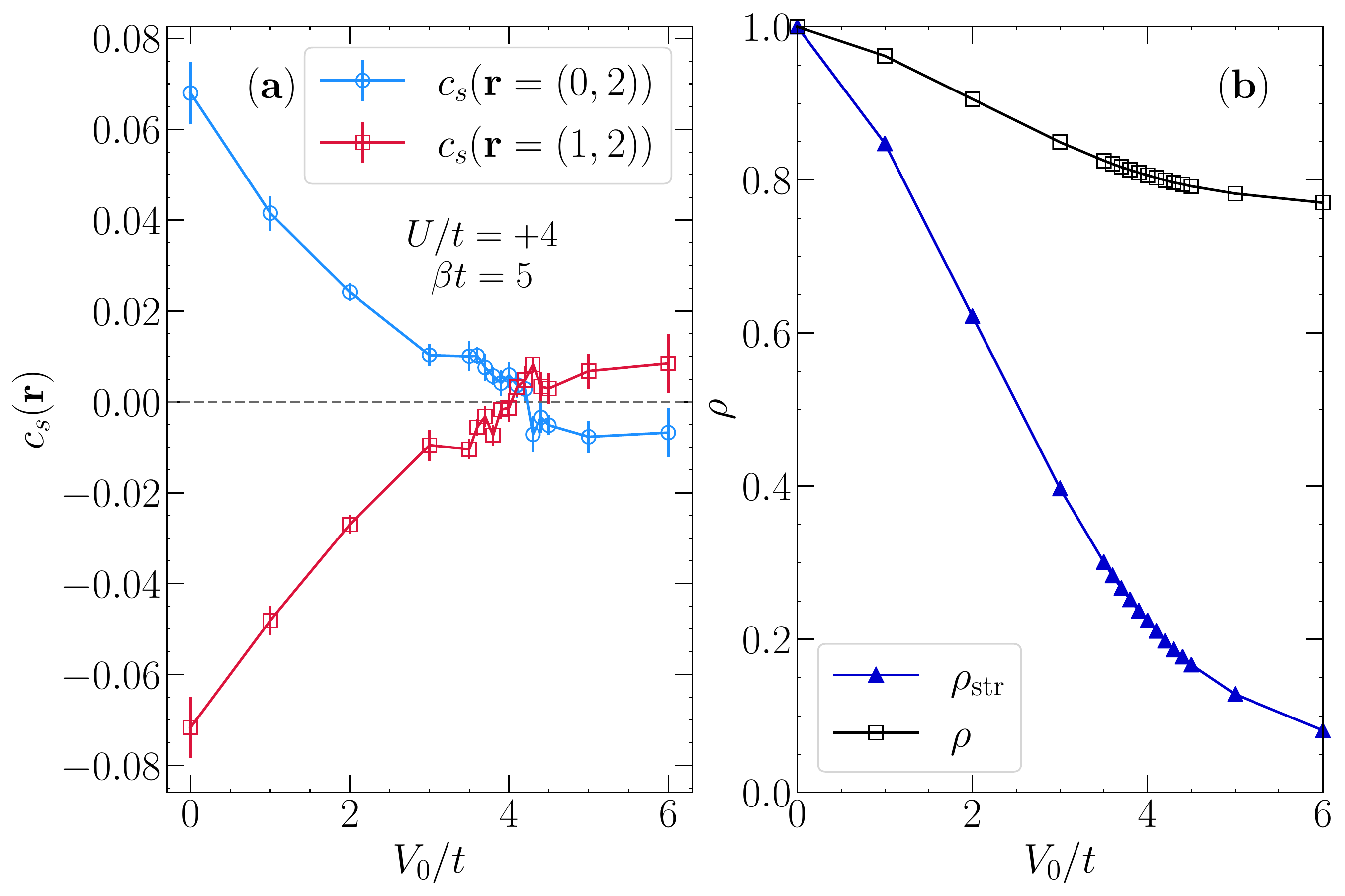}
\caption{(a) Spin-spin correlator 
$c_s\big({\bf r} = (0,2)\big)$ and 
$c_s\big({\bf r} = (1,2)\big)$, 
as functions of stripe strength $V_0$ for the \textit{repulsive} Hubbard model. \textit{Magnetic} $\pi$-phase shifts are formed at $V_0 \sim 4t$, where the reversal in sign of the spin correlation functions takes place. (b) The total electron density $\rho$ and the density on the stripe $\rho_{\rm str}$, as functions of stripe strength $V_0$. As in Fig.~\ref{corr_cross_rhooff1}, the off-stripe electron density, $\rho_{\rm dom}$, is fixed to be $1$ by adjusting the global chemical potential $\mu$. Here the lattice size is $8 \, \times \, 8$, the on-site repulsive interactions is $U = +4\,t$, and the inverse temperature $\beta \,t = 5$.
}
\label{fig:app_repulsive}
\end{figure}

A direct parallel to the results in the main text is given by the repulsive Hubbard model ($U>0$). These can display a magnetization reversal across a hole-rich region, leading to a \textit{magnetic} $\pi$-phase shift~\cite{mondaini12}, much like what is observed in certain classes of cuprates with static stripe formation~\cite{tranquada95,Tranquada2004}. Figure \ref{fig:app_repulsive} gives the equivalent of Fig.~\ref{corr_cross_rhooff1}, but for $U/t = +4$ instead, and similarly tuning $\rho_{\rm dom} = 1$. Here, we display the spin correlations,
\begin{align}
c_s({\bf r}) 
=\big\langle \hat S_{\bf i}^z \hat S_{\bf i+r}^z \big\rangle,
\end{align}
connecting the same [different] sublattice $c_s\big({\bf r} = (0,2)\big)$ [$c\big({\bf r} = (1,2)\big)$] across a stripe line. Due to the presence of the sign problem, we focus instead at higher temperatures, $\beta t =5$. Although qualitatively similar, the value of the stripe energy that leads to the AF $\pi$-phase shift is much larger ($V_{0c}\simeq 4t$) than the one that leads to the density $\pi$-phase shift ($V_{0c}\simeq t$), both of which with $|U|/t=4$; 
yet, the stripe electronic filling at which the magnetization reversal takes place is similar, $\rho_{\rm str} \simeq 0.23$.

The fact that larger $V_{0}$ is required to lower the density on a stripe in the repulsive case is due to the presence of 
a Mott gap.  Single occupancy of sites is energetically preferred, and $V_0$ must overcome that tendency.  
Indeed,  when thermal
and quantum fluctuations are turned off, $T=t=0$,
$V_{0c} = U$ is precisely the critical value for stripe formation. This can be seen by considering a two
site system with $V_0$ on one site:  
The configuration $| \uparrow \,\,\,\, \downarrow \,\,\rangle$ with two singly occupied sites  has energy 
$-\frac{U}{4} + \big(-\frac{U}{4}+V_0\big)$.
The configuration $| \uparrow \downarrow \,\,\,\, * \,\,\rangle$ with a doubly occupied and an empty site  has energy 
$+\frac{U}{4} + \frac{U}{4}$.  These become degenerate at $V_{0c} = U$, in agreement with 
Fig.~\ref{fig:app_repulsive}.

Meanwhile, for the attractive case, double occupation is already favored (on alternating sites).  A potential $V_0$ on
a linear set of sites needs only to overcome the charge alternation, an energy scale $\sim 4 t^2/U$.
This suggests $V_{0c} \sim t$, again, in rough agreement with 
Fig.~\ref{corr_cross_rhooff1}.
In any case, the difference between the values of $V_{0c}$ stresses that the two models are \textit{not} connected by PHT.

\section{Trotter Errors} \label{sec:app_Trotter}
\begin{figure}[h] 
\includegraphics[width=0.99\columnwidth]{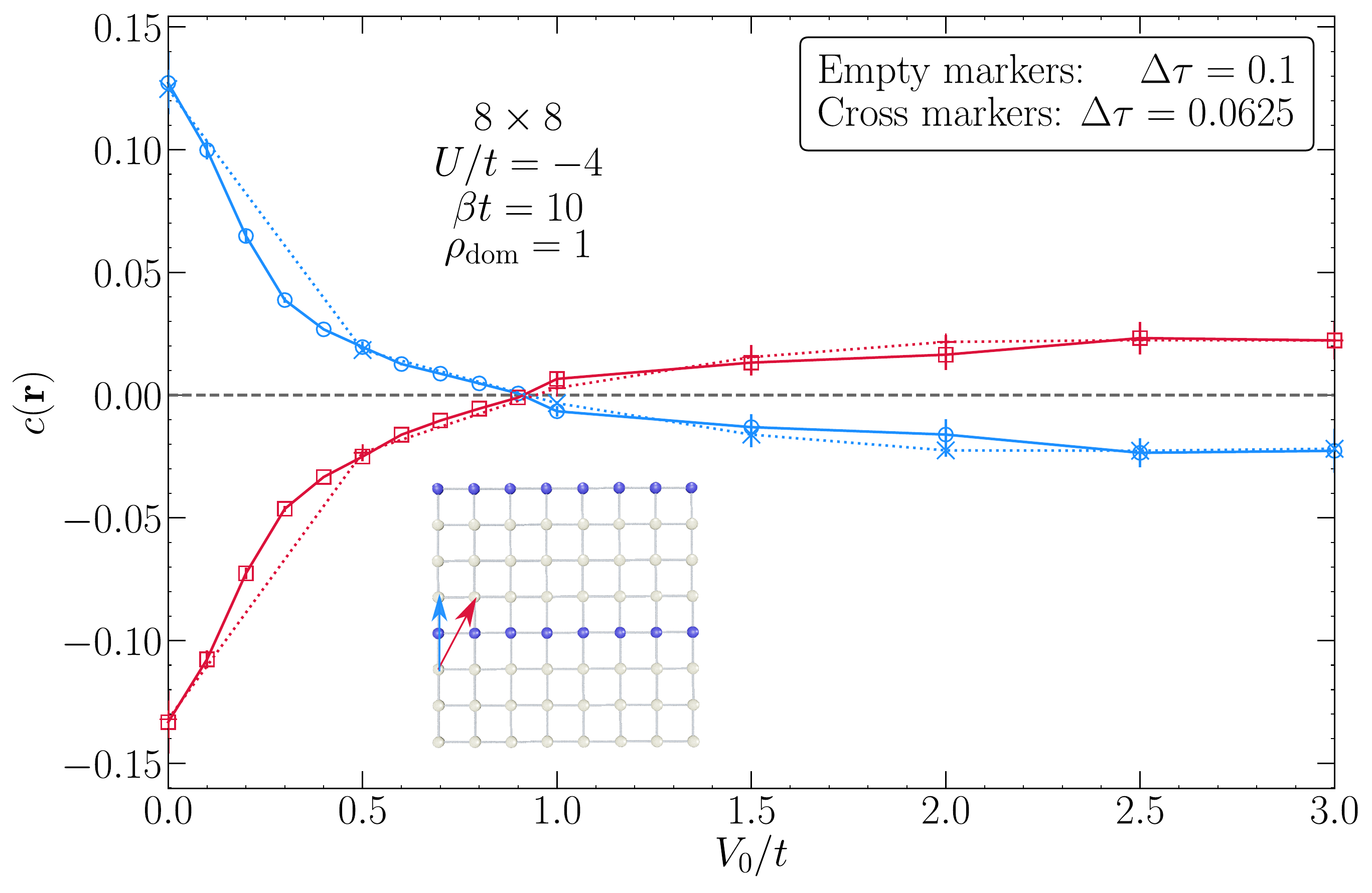}
\caption{Density-density correlator $c\big({\bf r} = (0,2)\big)$ (blue symbols and blue arrow in the inset) and $c\big({\bf r} = (1,2)\big)$ (red markers, and red arrow in the inset), as functions of stripe strength $V_0$, for two different imaginary time discretizations $\Delta \tau=0.1$ (empty markers), the value presented in the main text, and $\Delta \tau=0.0625$ (cross markers). The results are the same to within the statistical error bars.
}
\label{fig:app_Trotter}
\end{figure}

In Fig.~\ref{fig:app_Trotter}, we re-compute the results for $c({\bf r})$ of Fig.~\ref{corr_cross_rhooff1} at a smaller $\Delta \tau=0.0625$.  The values are not significantly shifted, and conclusions of the main text are unaltered.

\section{Finite temperature and finite size effects}  \label{sec:app_fse_fte}

\begin{figure}[h] 
\includegraphics[width=0.99\columnwidth]{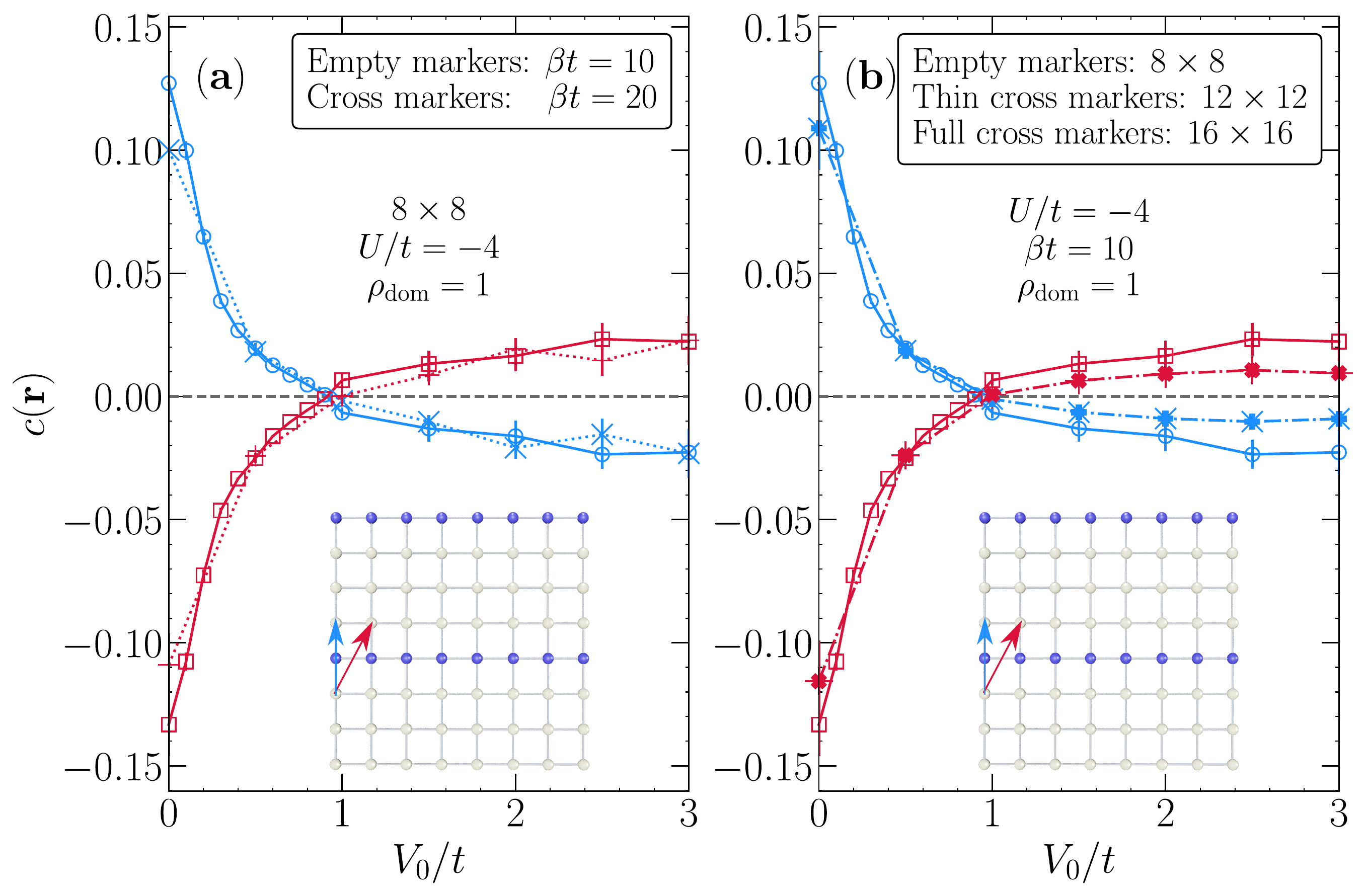}
\caption{(a) Density-density correlations 
$c\big({\bf r}=(0,2)\big)$ (blue markers, and blue arrow in the inset) and $c\big({\bf r}=(1,2)\big)$ (red markers, and red arrow in the inset), as functions of stripe strength $V_0$, for two different temperatures $\beta=10$ (empty markers), the value
presented in the main text, and $\beta=20$ (cross markers) for an $8 \times 8$ lattice. The results are the same to within the statistical error bars. (b) Same correlations as functions of stripe strength $V_0$, but now for three different lattice sizes $8 \times 8$ (empty markers), $12 \times 12$ (thin cross markers) and $16 \times 16$ (full cross markers) at $\beta t = 10$. Data parameters in both panels are $U/t = -4$ and $\rho_{\rm dom} =1$.
\label{fig:fse_fte}
}
\end{figure}

Here we explore the robustness of our results to lowering the temperature further and to increasing the lattice size. Figure \ref{fig:fse_fte}(a) compares the results for the density correlations across a stripe of Fig.~\ref{corr_cross_rhooff1}(a) (at $\beta\,t=10$) with DQMC simulations at $\beta\,t=20$. The $\pi$-phase shift is still observed, and occurs at the same $V_{0c}\sim t$.

To quantify finite size effects, we increase the lattice size to 
$12 \times 12$,
leaving the other parameters unchanged. The $\pi$-phase shift is still observed,
as shown in Fig.~\ref{fig:fse_fte}(b).
There is some reduction in the density correlations in going from
$8 \times 8$ to
$12 \times 12$ at large $V_0$.  However, the
$16 \times 16$ lattice results lie on top of those for
$12 \times 12$,
indicating convergence to a non-zero $\pi$-phase shifted value
in the thermodynamic limit.

\section{Results for general $\rho_{\rm dom}$} \label{sec:app_other_dens}

\begin{figure}[h]
\includegraphics[width=0.99\columnwidth]{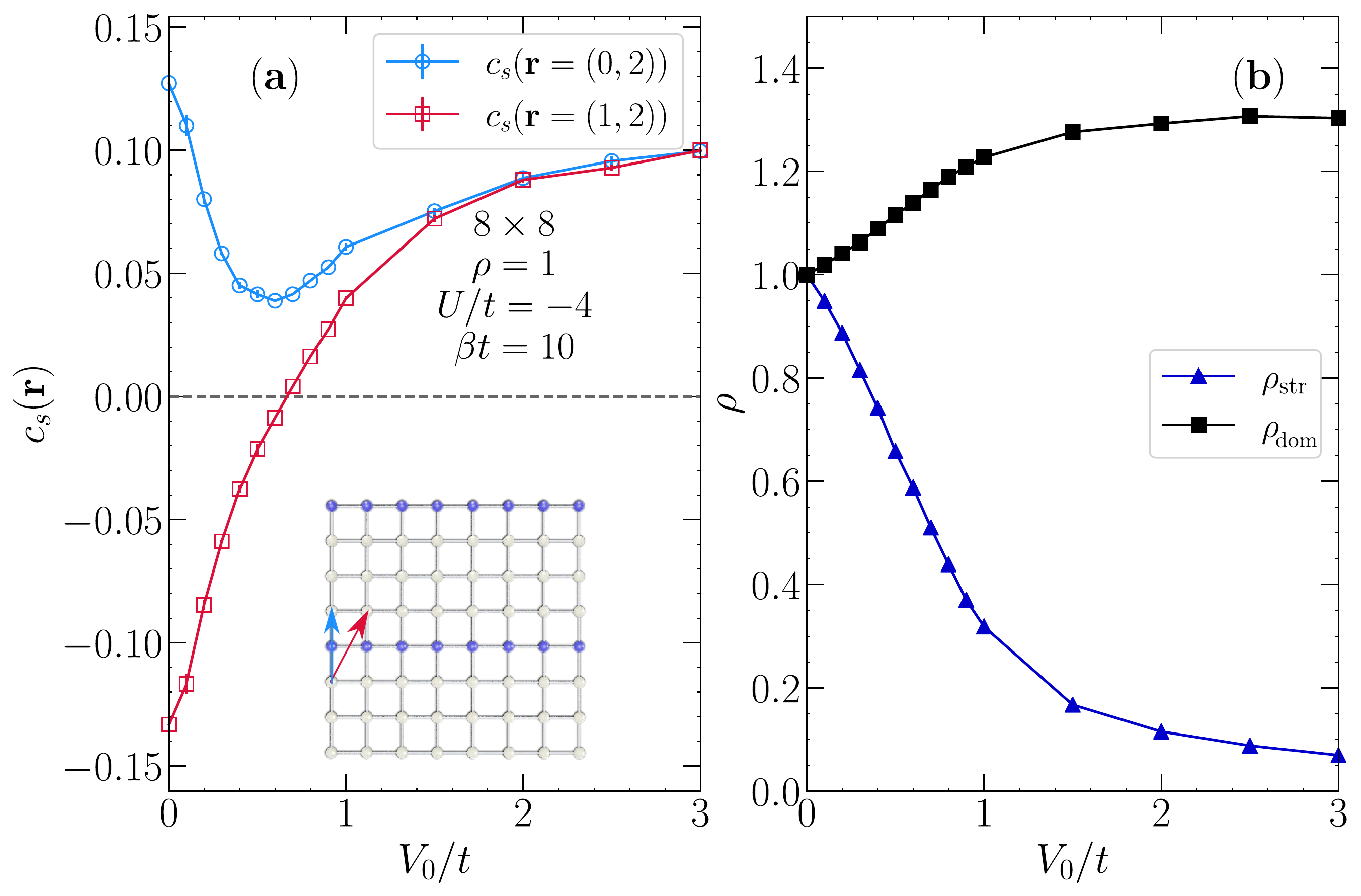}
\caption{(a) Similar to Fig.~\ref{corr_cross_rhooff1}(a), but with the 
{\it total} electron density 
$\rho$ fixed to be $1$. Although 
$c\big({\bf r}=(1,2)\big)$ exhibits a sign change,
$c\big({\bf r}=(0,2)\big)$ does not.
When the interstripe domains are not pinned at half-filling, complete
sublattice reversal across a stripe does not occur. (b) The electron density on and off the stripe, as functions of stripe strength $V_0$, with the total electron density fixed to be $1$, and for parameters listed in the figure.
\label{fig:rhototal1}
}
\end{figure}

In Fig.~\ref{fig:rhototal1}(a), we show two density-density correlations  
traversing the stripe 
$c\big({\bf r}=(1,2)\big)$ and
$c\big({\bf r}=(0,2)\big)$ 
for simulations in which $\mu$ is not tuned to keep $\rho_{\rm dom}=1$ but instead
the total density $\rho=1$.
Since $\rho=\frac{1}{4} \rho_{\rm str}
+\frac{3}{4} \rho_{\rm dom}$
and $\rho_{\rm str} \rightarrow 0$ as $V_0$ becomes large, the density
in the inter-stripe domains interpolates between 
$\rho_{\rm dom} = 1$
and
$\rho_{\rm dom} = \frac{4}{3}$
as $V_0$ goes from $V_0=0$ to $V_0=\infty$.
This is seen in Fig.~\ref{fig:rhototal1}(b).
Although $c\big({\bf r}=(0,2)\big)$ initially decreases from its positive value at $V_0=0$, it recovers and always remains positive.  There is no clear signature of $\pi$-phase shift.

\begin{figure}[h]
\includegraphics[width=0.99\columnwidth]{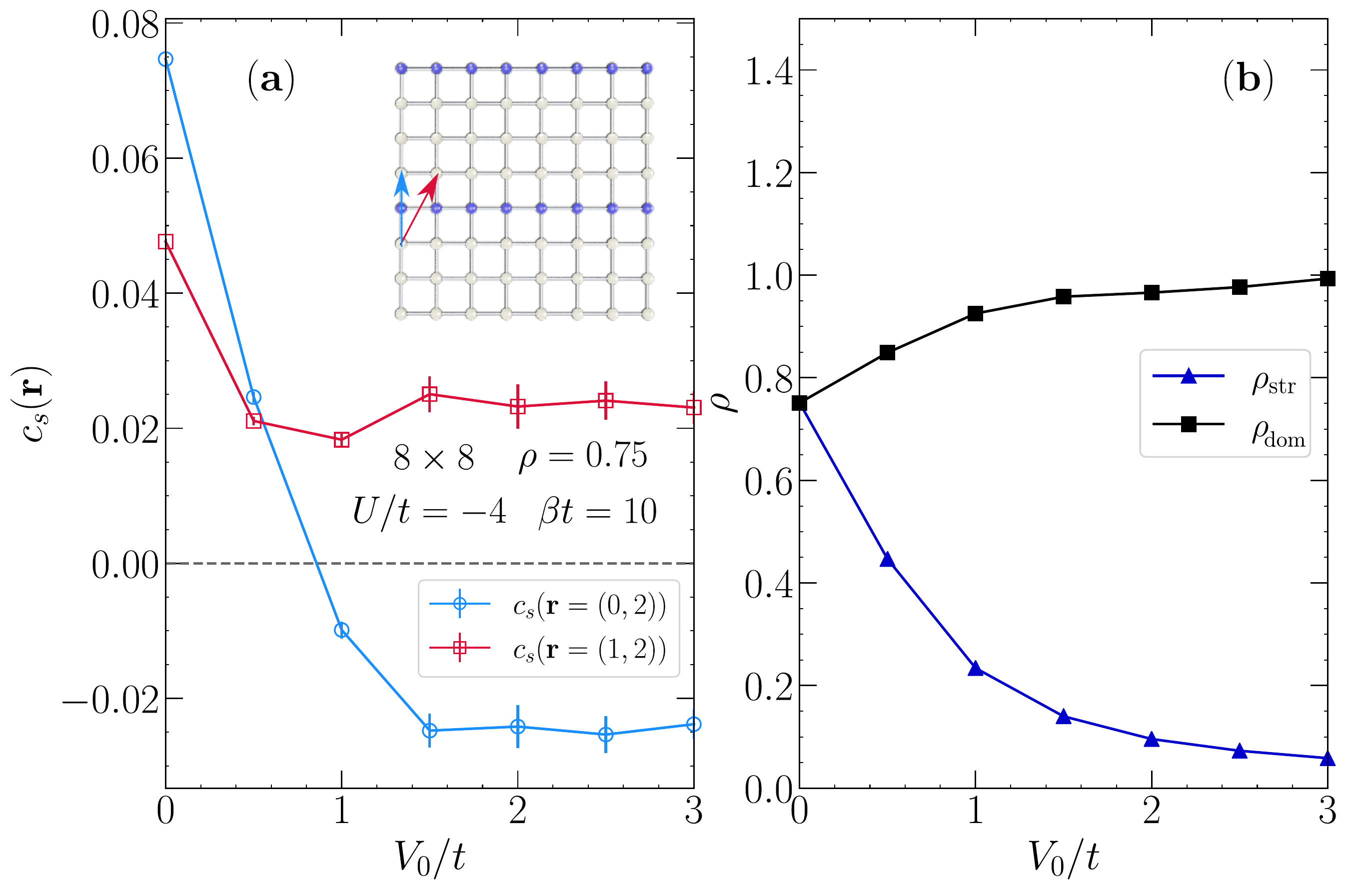}
\caption{Same as Fig.~\ref{fig:rhototal1} but with the total electron density fixed to be $0.75$, and for the parameters listed in the figure. 
\label{fig:rhototal075}
}
\end{figure}

We next performed QMC calculations fixing the total electron density to be $0.75$, as shown in Fig.~\ref{fig:rhototal075}(a). We find that, with the increase of $V_0$, $c\big({\bf r}=(0,2)\big)$ goes from positive to negative values as should occur for a $\pi$-phase shift. Meanwhile $c\big({\bf r}=(1,2)\big)$ {\it begins} already at $V_0=0$ with a positive value appropriate to a $\pi$ phase shift, and remains so as $V_0$ grows. Thus at large $V_0$ both correlation functions are consistent with sublattice reversal. The corresponding densities are shown in Fig.~\ref{fig:rhototal075}(b).

\begin{figure}[t!] 
\includegraphics[width=0.99\columnwidth]{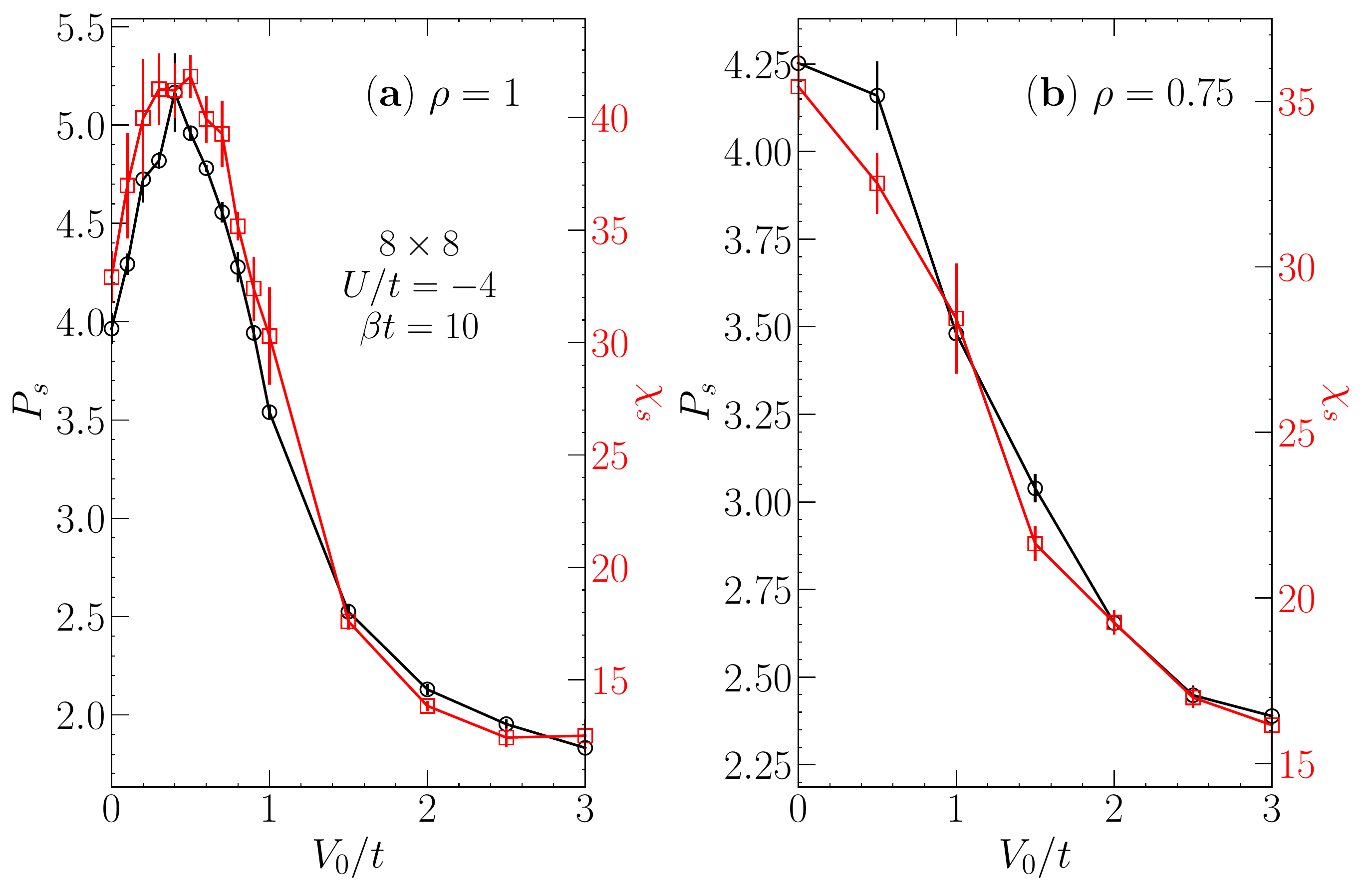}
\caption{The $s$-wave pairing structure factor $P_s$, left axis (pairing susceptibility $\chi_s$, right axis), as a function of $V_0$, similar to Fig.~\ref{Ps_chi_s}, but for $\rho = 1$ (a) and $\rho = 0.75$ (b).
\label{fig:app_Ps_chis_other_dens}
}
\end{figure}

In Fig.~\ref{fig:app_Ps_chis_other_dens}, we show the $s$-wave pairing structure factor $P_s$ and the $s$-wave pairing susceptibility $\chi_s$, separately, as functions of $V_0$. The two panels are similar to Fig.~\ref{Ps_chi_s}, but for $\rho = 1$ and $\rho = 0.75$. For $\rho=1$ there are peaks in $P_s$ and $\chi_s$ similar to those  occurring for fixed $\rho_{\rm dom}=1$. 
However $\rho=0.75$ shows no enhancement of SC with the imposition of stripes.

\section{Exact diagonalization results} \label{sec:app_ed}

\begin{figure}[h] 
\includegraphics[width=8.0cm,keepaspectratio]{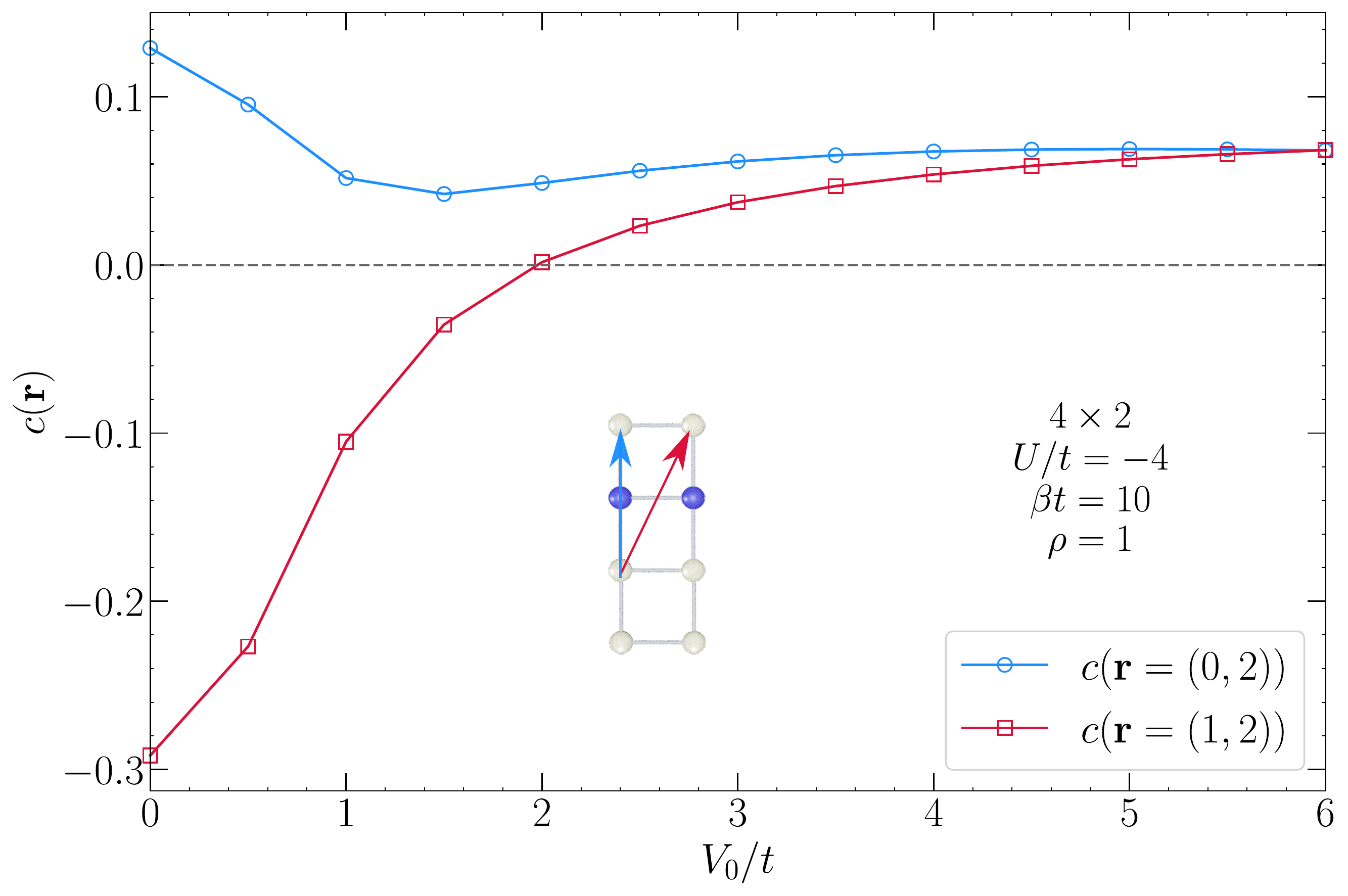}
\caption{ED results of the density-density correlator 
$c\big({\bf r} = (0,2)\big)$
(blue circles and
blue arrow in the inset) and 
$c\big({\bf r} = (1,2)\big)$
 (red squares, and red arrow in the inset), 
 as functions of stripe
 strength $V_0$.  The total electron density is fixed to be $1$. 
\label{fig:corr_cross_rhototal1-ED}
}
\end{figure}

As a complement to the QMC data investigated, we also performed ED calculations on a $2 \times 4$ lattice, with the total electron density fixed to be $1$, as shown in Fig.~\ref{fig:corr_cross_rhototal1-ED}. 
The density correlators $c\big({\bf r}=(0,2)\big)$ 
and $c\big({\bf r}=(1,2)\big)$ 
behave in a qualitatively similar way to the QMC data in Fig.~\ref{fig:rhototal1}(a) of Appendix \ref{sec:app_other_dens}.

\bibliography{posustripes}

\clearpage

\renewcommand{\thefigure}{S\arabic{figure}}
\setcounter{figure}{0}
\renewcommand{\thesection}{S\arabic{section}}
\setcounter{section}{0}
\renewcommand{\theequation}{S\arabic{equation}}
\setcounter{equation}{0}

\end{document}